\documentstyle[12pt]{article}
\begin{document}
\newcommand{\be}{\begin{equation}}
\newcommand{\ee}{\end{equation}}
\begin{titlepage}
\title{Effects of the gravivector and graviscalar fields \\ in $N=2,8$
supergravity$^{*}$}
\author{Stefano Bellucci \\ 
           \\{\small \it INFN--Laboratori Nazionali di Frascati, 
P.O. Box 13, I-00044 Frascati, Roma (Italy)}  
\\ \\ and \\ \\
Valerio Faraoni \\ 
        \\{\small \it Department of Physics and Astronomy, University 
of Victoria} \\
{\small \it P.O. Box 3055, Victoria, B.C. V8W 3P6 (Canada)}}
\date{}
\maketitle
\thispagestyle{empty}      \vspace*{1truecm}
\begin{abstract}
The available tests of the equivalence principle constrain the mass of
the Higgs-like boson appearing in extended supergravity theories.
We determine the constraints imposed by high
precision experiments on the antigravity fields (gravivector and graviscalar) 
arising from $N=2,8$ supergravity.
\end{abstract} 
                    \vspace*{1truecm}
* Accepted for publication in Phys. Lett. B. - PACS: 04.65.+e 04.80.Cc
\vfill
\begin{center}
May 1996
\end{center}
\end{titlepage}  

The discovery that $N>1$ supergravity theories lead to antigravity
is due to the work of the late J. Scherk \cite{Scherk,ScherkProc}.
In a recent paper we have revived the interest for the implications
of extended supergravity theories for antigravity \cite{belfar}.
This interest is connected to the high precision experiment at LEAR
(CERN) measuring the difference in the gravitational acceleration of 
the proton and the antiproton \cite{PS-200}. For a review of earlier ideas 
about antigravity
the reader is referred to the extensive article by Nieto and
Goldman \cite{GoldmanNieto} and the references therein.
 
The supergravity multiplet in the $N=2,8$ cases contains, in addition to
the graviton ($J=2$), a vector field $A_{\mu}^l$ ($J=1$). There 
are also two Majorana gravitini ($J=\frac{3}{2}$) for $N=2$ \cite{Zachos}
and a scalar field $\sigma$ for $N=8$ \cite{Scherk,ScherkProc}.
The former fields are immaterial for our purposes and will be ignored in 
the following. It is also to be noted that there are
important differences between extended supergravity and the Standard Model,
and therefore the particles mentioned in this work should not be intended
as the objects familiar from the Standard Model.

The field $\sigma$ (called graviscalar in what follows) introduces a violation
of the equivalence principle in the form of a universal (i.e. independent from
the composition of the material) spatial dependence in Newton's constant,
$G=G(r)$. However, this violation does not affect any E\"{o}tvos--like
experiment measuring differences in the acceleration of bodies of different
composition. Hence, the only way to constrain the effective range of the
interaction mediated by the $\sigma$--field is by means of experiments testing
deviations from Newton's law such as those searching for a fifth force. In
contrast, the effect of the gravivector $A_{\mu}^{l}$ depends on the
composition of test bodies, and is most effectively constrained by
E\"{o}tvos--like experiments.

The E\"otv\"os experiment forces upon us the assumption that the field
$A_{\mu}^l$ have a nonvanishing mass, which may have a dynamical origin
\cite{Scherk,ScherkProc}. In any case, the vector receives a mass
through the Higgs mechanism\setcounter{equation}{0}
\be                          \label{1}
m_l=\frac{1}{R_l}=k\, m_{\phi}\langle \phi \rangle \; ,       
\ee
where $k=(4\pi G)^{1/2}$ and the mass of the Higgs--like 
field equals its (nonvanishing) vacuum expectation value ({\em v.e.v.})
\be                          \label{2}
m_{\phi}=\langle \phi \rangle \; .       
\ee
Thus, Scherk's model of antigravity leads to the possibility of
violating the equivalence principle on a range of distances of order
$R_l$, where $R_l$ is the $A_{\mu}^l$ Compton wavelength. The available
limits set by the experimental tests of the equivalence principle
allow us to constrain the {\em v.e.v.} of the Higgs-like field $\phi$, and
therefore its mass. It must be noted that the possibility of a 
massless field $A_{\mu}^l$ was already ruled out by Scherk using the 
E\"{o}tv\"{o}s experiments available at that time \cite{Scherk}.

In the present paper we build upon \cite{belfar}: taking into account the
experiments up to date, we are able to improve the limits on the gravivector
$A_{\mu}^l$. Moreover, we extend our treatment by
considering the effects of the graviscalar for the case $N=8$, and provide
the constraints set by fifth force experiments (non E\"{o}tvos--like tests
of the equivalence principle) and by the binary pulsar PSR~1913+16.

The Compton wavelength of 
the gravivector already obtained in \cite{belfar} is of order 10~m, or 
less. Incidentally, the smallness of this upper bound justifies the
use of E\"{o}tvos--like experimental results, which lose their
validity at much larger distances. Therefore, 
the concept of antigravity in the context of $N=2$ supergravity 
cannot play any role in astrophysics, except possibly for processes 
involving the strong gravity regime, i.e. near black holes or in the 
early universe. The same conclusion applies to the case $N=8$, owing to the
results we present here, since the $N=8$ graviscalar and gravivector effective
ranges of interaction are constrained, respectively, to be less than 100~m and
1~m.

A caveat concerning our results for the graviscalar is worth mentioning: our
analysis and conclusions for the interaction of this field with matter and
antimatter are by no means exhaustive, and our experimental limits hold only
for the field $\sigma$ entering the $N=8$ supergravity multiplet. For a
treatment of the couplings of a Brans--Dicke scalar in various other models,
we refer the reader to
\cite{GoldmanNieto}. Alternatively, ultra--light pseudo Nambu--Goldstone
bosons have been considered in extensions of the standard model
\cite{HillRoss} and
observational constraints based on astrophysical considerations have been
obtained \cite{FriemanGradwohl}.

In $N=2$,~$8$ supergravity theories, the gravivector field $A_{\mu}^l$ 
couples to the fields of the matter scalar multiplet with strengths
\be   \label{3}
g_i=\pm k\, m_i                  \ee
\cite{Zachos} for $N=2$ and 
\be            
g_i=\pm 2k\,m_i                     \ee
\cite{SS,CremmerSS} for $N=8$. Here $m_i$ are the 
quark and lepton masses, the positive and negative signs hold for 
particles and antiparticles, respectively, and $g=0$ for 
self--conjugated particles. As a consequence, in the interaction of an 
atom with the gravitational field, the vector field $A_{\mu}^l$ 
``sees'' only the particles constituting the nucleon which are not 
self--conjugated, while the graviton and the graviscalar (for $N=8$) couple 
to the real mass of the nucleon. 

For two composite particles, e.g. two atoms
with masses $M_1$, $M_2$ at separation $r$, the potential energy reads 
\be      \label{atoms}
V(r)=-\,\frac{GM_1 M_2}{r} \left[ 1+ \alpha_l \exp(-r/R_l) 
+\alpha_{\sigma} \exp(-r/R_{\sigma}) \right]  \; , \ee
where 
\be \label{alfa} 
\alpha_l = -\,\frac{M_1^0 M_2^0}{M_1 M_2}\,  \eta\:, \;\;\;\;\;\;\;\;
\alpha_{\sigma} =\eta -1 \; ,\;\;\;\;\;\;\;\;
\eta =\left\{ \begin{array}{cllll}
1 & \,\,\,\, , & \;\; N=2  &  & \nonumber \\
4 & \,\,\,\, , & \;\; N=8  &  & \nonumber 
\end{array} \right. 
\; 
\ee
and $R_l$ ($R_{\sigma}$) is the Compton wavelength of the gravivector
(graviscalar).
The masses in (\ref{atoms}), (\ref{alfa}) are given by
\begin{eqnarray}
&& M=Z(M_p+m_e)+(A-Z)M_n \; ,   \label{7}  \\ 
&& M^0=Z(2m_u+m_d+m_e)+(A-Z)(m_u+2m_d) \; ,  \label{8}
\end{eqnarray}
where $Z$ and $A$ are the atomic and
mass numbers and $M_p$, $M_n$, $m_e$, $m_u$ and $m_d$ are the proton, 
neutron, electron, up quark and down quark masses, respectively. We use the
values $m_u=5.6$~MeV,
$m_d=9.9$~MeV, consistently with \cite{belfar}. Notice that in the case 
$N=8$, $\alpha_{\sigma}$ is three orders of magnitude larger than
$\alpha_l$. 
In fact, substituting the values of the masses in eqs.~(\ref{7}), (\ref{8})
one obtains
\be
\frac{M^0}{M}=\frac{-3.8 Z+25.4A}{-0.8Z+939.6A} \leq \xi \; ,
\ee
where $\xi=2.7 \cdot 10^{-2}$, and the inequality $A\geq Z$ has been used.
Hence, we have $|\alpha_l|\leq \eta \xi^2$ which, for $N=8$, yields the limit
$|\alpha_l|\leq 2.9 \cdot 10^{-3}$.

We consider high--precision tests of the equivalence principle and its
violation induced by antigravity in $N=2,8$ supergravity, in order to get
observational bounds on the effective range of the vector gravity interaction
and the Higgs--like boson appearing in the theory \cite{belfar}. The sign and
the strength of the coupling of the graviscalar $\sigma$ is the same for all
particles and antiparticles.
Since the coupling of the graviscalar is universal, the contribution of
spin~0 gravity to the acceleration of a test body does not depend on its
composition.
Therefore, this contribution does not affect the difference $\delta \gamma$
of the gravitational accelerations of two test bodies with different
compositions.
When considering E\"{o}tvos--like experiments, it is safe to omit the scalar
$\sigma$, and the potential for an atom in the static field of the Earth 
is \cite{Scherk}
\be      \label{4}
V=-\,\frac{G}{r} \left[ MM_{\oplus}-\eta M^0 {M^0_{\oplus}} \, 
f\left( \frac{R_\oplus}{R_l} \right) \exp(-r/R_l) \right]  \; , \ee
where $R_\oplus =6.38 \cdot 10^6$~m and $ M_{\oplus}=5.98 \cdot 10^{24}$~kg 
are the earth radius and mass, respectively. The presence of the 
function 
\be     \label{6}
f(x)=3 \,\, \frac{x\cosh x-\sinh x}{x^3}      \ee
expresses the fact that a spherical mass distribution cannot be 
described by a point mass located at the 
center of the sphere, as in the case of a coulombic potential. We describe 
the Earth by means of the average atomic composition 
$(Z_{\oplus},2Z_{\oplus})$ which gives, from (\ref{7}),~(\ref{8})
\be   \label{9}
M^0_{\oplus} \simeq \frac{3m_u+3m_d+m_e}{M_p+M_n} \, M_{\oplus} \; .
\ee

In $N=2,8$ supergravities, one of the scalar fields (other than $\sigma$) 
has a nonzero 
{\em v.e.v.} and, as a consequence, the vector field $A_{\mu}^l$ acquires a 
mass, as described by (\ref{2}) (the impossibility of a massless 
$A_{\mu}^l$ being proved in ref.~\cite{Scherk}). This leads 
to a violation of the equivalence principle, expressed by the 
difference between the accelerations of two atoms with numbers 
$(Z,A)$ and $(Z',A')$ in the field of the Earth
\be            \label{10}                           
\frac{\delta \gamma}{\gamma}=\eta \, \frac{(3m_u+3m_d+m_e)(m_e+m_u-m_d)}
{M_n \, (M_p+M_n) } \left( \frac{Z'}{A'}-\frac{Z}{A} \right) f\left( 
\frac{R_{\oplus}}{R_l}\right) \left( 1+\frac{R_{\oplus}}{R_l} \right) 
\exp(-R_{\oplus}/R_l) \; .                    \ee
In the E\"{o}tv\"{o}s--like experiment performed at the University of 
Washington \cite{EotWash} (hereafter ``E\"{o}t--Wash'') the equivalence 
principle was tested using berillium and copper and aluminum and 
copper. This test
was used in ref.~\cite{belfar} to set a lower limit on the mass 
of the Higgs--like particle
\be    \label{12}
m_{\phi}>5 \,  \eta^{1/2} \;\;\;\mbox{GeV}        \; .    \ee

The E\"{o}t-Wash experiment has recently been improved \cite{Suetal},
yielding the higher precision limit
\be     
\left| \frac{\delta \gamma}{\gamma} \right| \leq 3.0 \cdot 10^{-12} 
\ee
for berillium and aluminum, which translates into the improved upper
limit for the gravivector
\be     \label{newRl}
R_l \leq 3.4 \, \eta^{-1} \,\,\,\,\, \mbox{m}
\ee
or equivalently,
\be
m_{\phi} \geq 15.8 \, \eta^{1/2}     \;\;\;\mbox{GeV}        \; .    
\ee
It is also to be noted that by increasing the factor 
$ \left( \frac{Z'}{A'}-\frac{Z}{A} \right)$ in~(\ref{10}), the 
upper limit on $R_l$ can be
improved. This was achieved in the last version of the E\"{o}t--Wash
experiment, where the best limit comes from the use of
berillium--aluminum 
($ \frac{Z'}{A'}-\frac{Z}{A}=0.038$) instead of berillium--copper 
($ \frac{Z'}{A'}-\frac{Z}{A}=0.012$) or aluminum--copper 
($ \frac{Z'}{A'}-\frac{Z}{A}=0.025$), which were used in the latest and 
in previous versions of the experiment.

We also consider the experiments aimed to detect deviations 
from Newton's inverse square law. In these experiments it is customary to
parametrize the deviations from the Newtonian form with a Yukawa--like 
correction to the Newtonian potential
\be   \label{13}
V(r)=-\,\frac{GM}{r} \left( 1+\alpha \, \mbox{e}^{-r/R_l} \right) \; .
\ee
In the following, we assume that, in the context 
of antigravity, the parameter $\alpha$  is given by the value computed
for the E\"{o}t--Wash 
experiment performed using copper ($Z=29 $, $A=63.5 $) and berillium
($Z'=4 $, $A'=9.0  $), i.e.
\be  \label{14}
\alpha =\left\{ \begin{array}{cllll}
6.36 \cdot 10^{-4} & \,\,\,\,\,\,  & \;\;\; (N=2)  &  & \nonumber \\
2.54 \cdot 10^{-3} & \,\,\,\,\,\,  & \;\;\; (N=8) \;.  &  & \nonumber 
\end{array} \right.    
\ee
For the materials that are likely to be used in these experiments, the 
values of $\alpha$ differ from those of (\ref{14}) only for a 
factor of order unity. Moreover, our final limits on $m_{\phi}$ depend 
on the square root of $\alpha$. For these reasons, it is safe to use 
the values (\ref{14}) of $\alpha$ in the following computations (it is to
be remarked that all the experiments considered in what follows measure
the gravitational attraction between bodies in a laboratory).

Equations (\ref{1}) and (\ref{2}) provide us with the relation
\be
\frac{m_{\phi}( \mbox{new})}{{m_{\phi}}^*}=\left( 
\frac{{R_l}^*}{R_l( \mbox{new})}\right)^{1/2} \; ,
\ee
where ${m_{\phi}}^*=5 \eta^{1/2}$~GeV and ${R_l}^*=34 \eta^{-1}$~m are,
respectively, the 
lower limit on the scalar field mass and the 
upper limit on the Compton wavelength 
of the vector $A_{\mu}^l$ derived in ref.~\cite{belfar}, and 
$m_{\phi}( \mbox{new})$, $R_l( \mbox{new})$ are the new limits on the 
same quantities coming from the references considered in the following.

The 2$\sigma$ limits of ref.~\cite{Spero} (see their fig.~3) allow the 
range of values of $R_l$:
\be              \label{15}
R_l \leq 1 \: \mbox{cm} \:\:\:\: , \:\:\:\: R_l\geq 5 \: \mbox{cm} 
\ee
for $N=2$ and
\be           \label{16}
R_l \leq 0.5 \: \mbox{cm} \:\:\:\: , \:\:\:\: R_l\geq 16 \: \mbox{cm} 
\ee
for $N=8$. This corresponds to the allowed range for the mass of the 
Higgs--like scalar field: 
\be    \label{17}
m_{\phi} \leq 130 \: \mbox{GeV} \:\:\:\: , \:\:\:\: m_{\phi}\geq 292 
\: \mbox{GeV} \:\:\:\:\:\:\:\: (N=2) 
\ee
\be   \label{18}
m_{\phi} \leq 73 \: \mbox{GeV} \:\:\:\: , \:\:\:\:\: m_{\phi}\geq 412 \: 
\mbox{GeV} \:\:\:\:\:\:\:\: (N=8) \; .
\ee
The curve~A of fig.~13 in ref.~\cite{Hoskinsetal} gives
\be              
R_l \leq 0.6 \: \mbox{cm} \:\:\:\: , \:\:\:\: R_l\geq 10 \: \mbox{cm} 
\ee
for $N=2$ and
\be           
R_l \leq 0.4 \: \mbox{cm} \:\:\:\: , \:\:\:\: R_l\geq 32 \: \mbox{cm} 
\ee
for $N=8$. Equivalently,
\be
m_{\phi} \leq 92 \: \mbox{GeV} \:\:\:\: , \:\:\:\: m_{\phi}\geq 376
\: \mbox{GeV} \:\:\:\:\:\:\:\: (N=2) 
\ee
\be   
m_{\phi} \leq 52 \: \mbox{GeV} \:\:\:\: , \:\:\:\: m_{\phi}\geq 461 \: 
\mbox{GeV} \:\:\:\:\:\:\:\: (N=8) \; .
\ee
The null result of the Shternberg \cite{Shternberg} experiment reviewed
by Milyukov \cite{Milyukov} in the light of Scherk's work provides us 
with the limits:
\be              
R_l \leq 4 \: \mbox{cm} \:\:\:\: , \:\:\:\: R_l\geq 13 \: \mbox{cm} 
\ee
for $N=2$ and
\be           
R_l \leq 2.2 \: \mbox{cm} \:\:\:\: , \:\:\:\: R_l\geq 40 \: \mbox{cm} 
\ee
for $N=8$. These are equivalent to:
\be   
m_{\phi} \leq 82 \: \mbox{GeV} \:\:\:\: , \:\:\:\: m_{\phi}\geq 146 \: 
\mbox{GeV} \:\:\:\:\:\:\:\: (N=2) 
\ee
\be   
m_{\phi} \leq 46 \: \mbox{GeV} \:\:\:\: , \:\:\:\: m_{\phi}\geq 197 \: 
\mbox{GeV} \:\:\:\:\:\:\:\: (N=8) \; .
\ee
Therefore, the best available limits on the mass of the scalar field 
are given by
\be              \label{19}
m_{\phi} \leq 82 \: \mbox{GeV} \:\:\:\: , \:\:\:\: m_{\phi}\geq 
376 \: \mbox{GeV} \:\:\:\:\:\:\:\: (N=2) 
\ee
\be    \label{20}
m_{\phi} \leq 46 \: \mbox{GeV} \:\:\:\: , \:\:\:\: m_{\phi}\geq 461 \: 
\mbox{GeV} \:\:\:\:\:\:\:\: (N=8) \; .
\ee

The experiments analyzed above also constrain the range of the
graviscalar interaction for $N=8$. The deviation from pure spin~2 gravity
introduced by the gravivector and the graviscalar can be described by
introducing the effective gravitational ``constant'' \cite{ScherkProc}
\be  \label{G}
G_{eff}(r)=G\left[ 1+\alpha_l \left( 1+\frac{r}{R_l} \right) \exp(
-\frac{r}{R_l})+\alpha_{\sigma} \left( 1+\frac{r}{R_{\sigma}} \right) \exp(
-\frac{r}{R_{\sigma}}) \right] \; ,
\ee
where $\alpha_l$ and $\alpha_{\sigma}$ are given by~(\ref{alfa}). Notice
that
$\alpha_{\sigma}$ is a universal coupling constant, while $\alpha_l$ 
depends on the composition of test bodies.
The binary pulsar PSR~1913+16 \cite{PSR} can be used to constrain the range
of the graviscalar. The upper limit (\ref{newRl}) on the range of the
gravivector prevents it from affecting the dynamics of the binary pulsar. 
The emission of gravitational waves from the binary occurs due 
to the coherent motion of mass
distributions (the two neutron stars orbiting around each other) on the
scale
$a=1.4 \cdot 10^9$~m~ (the major axis of the binary \cite{PSR}), where 
$a>>R_l$. In the case $N=8$, if the range of the
graviscalar is $R_{\sigma} >> a$, one has for the binary pulsar  
$r\approx a$ and  $\exp( -r/R_{\sigma}) \approx 1 $ in~(\ref{G}). Under 
these assumptions, the analysis of ref.~\cite{FordHegyi} 
can be applied (see also \cite{GoldmanNieto}). In order for the 
observed orbital decay of the binary pulsar to
agree with the theory, it must be $\alpha_{\sigma}< 3 \cdot 10^{-3}$
\cite{FordHegyi}. This is clearly incompatible with the prescription
$\alpha_{\sigma}=3$ of $N=8$ supergravity and therefore, the range 
$R_{\sigma} >>
1.4 \cdot 10^9$~m for the graviscalar interaction is forbidden by the
binary
pulsar observations. The case $R_{\sigma} \approx a$ is excluded as well
using
the data from the Earth--Lageos--lunar experiments summarized in
fig.~1{\em a}
of ref.~\cite{TalFisch}. The experimental constraint in this range is
$\alpha_{\sigma} < 10^{-6}$, which is again incompatible with the
prediction
of $N=8$ supergravity.

The Shternberg experiment \cite{Shternberg} provides 
us with the limits on the range of the graviscalar:
\be   \label{Shter}
R_{\sigma} \leq 0.8 \: \mbox{cm} \:\:\:\: , \:\:\:\: 
R_{\sigma}\geq 14 \: \mbox{m} \;    . \ee
By combining the data of the Shternberg and the other experiments
reviewed in
\cite{Milyukov} one improves the limits (\ref{Shter}) as 
\be    
R_{\sigma} \leq 0.15 \: \mbox{cm} \:\:\:\: , \:\:\:\: 
R_{\sigma} \geq 70  \: \mbox{m} \; .
\ee
However, part of this range is already forbidden by the PSR~1913+16
data.
The fifth force experiments reviewed in \cite{TalFisch} allow only 
the regions
\be
R_{\sigma} \leq 1 \: \mbox{cm} \:\:\:\: , \:\:\:\: 
60 \: \mbox{m} \leq R_{\sigma}\leq 100 \: \mbox{m}  \:\:\:\: , 
R_{\sigma} \geq 10^{14} \: \mbox{m} \; . 
\ee
The first of these limits is compatible with, but less stringent than 
the constraints set by the experiments in \cite{Milyukov}. The third
region is
forbidden by the observational data on the binary pulsar. 

As a conclusion, the best available limits on the range of the
graviscalar 
derived from the various experiments quoted above are
\be
R_{\sigma} \leq 0.15 \: \mbox{cm} \:\:\:\: , \:\:\:\: 
70 \: \mbox{m} \leq R_{\sigma}\leq 100 \: \mbox{m}  \; .
\ee
The graviscalar $\sigma$, like the gravivector, cannot play
any significant role in astrophysics, except possibly near black holes
or in
the early universe, when the size of the universe (or of primordial
structures) 
is comparable to, or less than $R_{\sigma}$.

\end{document}